\documentclass[aps,prl,superscriptaddress,twocolumn,preprintnumbers,amsmath,amssymb]{revtex4-1}
\usepackage{graphicx}
\usepackage{subfigure}
\usepackage{epsfig}
\usepackage{dcolumn}
\usepackage{bm}
\usepackage{ulem}
\usepackage{color}
\usepackage{multirow}
\usepackage{slashed}

\def\sgn{{\rm sgn}}

\def\be{\begin{equation}}       \def\ee{\end{equation}}
\def\bea{\begin{eqnarray}}      \def\eea{\end{eqnarray}}
\def\ba{\begin{array}}
\def\ea{\end{array}}
\def\bnum{\begin{enumerate} }
\def\enum{\end{enumerate}}

\def\nn{\nonumber}

\def\=>{\Rightarrow}
\def\>{\rightarrow}

\def\eye2{Fathbb{I}}

\def\Eq#1{Eq.~(\ref{#1})}
\def\fig#1{Fig.~\ref{#1}}

\renewcommand{\>}{\rangle}

\begin{document}

\title{Solvable Sachdev-Ye-Kitaev models in higher dimensions: from diffusion to many-body localization}

\author{Shao-Kai Jian}
\affiliation{Institute for Advanced Study, Tsinghua University, Beijing 100084, China}

\author{Hong Yao}
\email{yaohong@tsinghua.edu.cn}
\affiliation{Institute for Advanced Study, Tsinghua University, Beijing 100084, China}
\affiliation{State Key Laboratory of Low Dimensional Quantum Physics, Tsinghua University, Beijing 100084, China}
\affiliation{Collaborative Innovation Center of Quantum Matter, Beijing 100084, China}

\begin{abstract}
Many aspects of many-body localization (MBL) transitions remain elusive so far. Here, we propose a higher-dimensional generalization of the Sachdev-Ye-Kitaev (SYK) model and show that it exhibits a MBL transition. The model on a bipartite lattice has $N$ Majorana fermions with SYK interactions on each site of the $A$ sublattice and $M$ free Majorana fermions on each site the of $B$ sublattice, where $N$ and $M$ are large and finite. For $r$$\equiv$$M/N\!<\!r_c$=1, it describes a diffusive metal exhibiting maximal chaos. Remarkably, its diffusive constant $D$ vanishes [$D$$\propto$$ (r_c-r)^{1/2}$] as $r$$\rightarrow$$r_c$, implying a dynamical transition to a MBL phase. It is further supported by numerical calculations of level statistics which changes from Wigner-Dyson ($r$$<$$r_c$) to Poisson ($r$$>$$r_c$) distributions. Note that no subdiffusive phase intervenes between diffusive and MBL phases. Moreover, the critical exponent $\nu$=0, violating the Harris criterion. Our higher-dimensional SYK model may provide a promising arena to explore exotic MBL transitions.
\end{abstract}
\date{\today}
\maketitle

As a foundation of equilibrium statistical mechanics, quantum thermalization and the eigenstate thermalization hypothesis (ETH) for closed quantum systems \cite{deutsch1991,srednicki1994,rigol2008} has received a surge of attention recently. A closed system satisfies ETH and acts as its own heat bath if interactions can thermalize all its own subsystems after long-time dynamic evolution. Nonetheless, it was conceived decades ago and was shown more recently that interacting systems with strong disorders may fail to thermalize but are many-body localized (MBL) \cite{anderson1980,giamarchi1987,altshuler1997,polyakov2005, altshuler2006,huse2007,huse2010,nayak2013, huse2015review, altman2015review,moore2016review}. Although properties of MBL phases are largely understood now \cite{sondhi2013prb,sondhi2014prb,demler2014prx,altman2014prl,abanin2013, huse2014prb,pollmann2012,cenke2016prb,sid2016arxiv,dassarma2017,xiechen2016,huizhai2016,swingle2016, fradkin2016,rqhe2016,yuchen2016}, dynamical phase transitions between MBL and thermal phases remain elusive despite the tremendous progress that has been achieved in understanding them \cite{lev2015,knap2016,varma2016,altman2015prx,sid2015prx}.

Here, we propose a solvable higher-dimensional generalization of the Sachdev-Ye-Kitaev (SYK) model \cite{kitaev2015,sachdev1993} and show that it features a dynamical phase transition between a thermal diffusive metal and a MBL phase.  The original SYK model consisting of large-$N$ disordered Majorana fermions in zero space dimension was proposed by Kitaev \cite{kitaev2015}, which is a generalization of the disordered spin model by Sachdev and Ye \cite{sachdev1993}. The SYK model is almost solvable in the large-$N$ limit with an approximate conformal or reparametrization symmetry in low temperature. Moreover, its Lyapunov exponent defined in out-of-time correlations (OTOC) \cite{larkin1969,kitaev2014, shenker2014} saturates the upper bound \cite{stanford2016b}, implying a holographic dual \cite{kitaev2015} to a dilaton gravity theory in nearly AdS$_2$ geometry \cite{stanford2016c, jensen2016, verlinde2016a}. Various aspects \cite{polchinski2016, you2016, yoon2016, stanford2016a, tezuka2016, verbaarschot2017, tanasa2017, gross2017, altland2017, franz2017, antonio2016} and interesting generalizations \cite{fu2016, altman2017, gu2016, gross2016, simon2016, sachdev2016, sachdev2017, witten2016, klebanov2017, volovich2016,verlinde2017, cenke2017, zhou2017, gu2017, krishnan2016, chen2017, song2017} of the SYK model have been studied so far.

\begin{figure}[t]
\subfigure{\includegraphics[width=6.5cm]{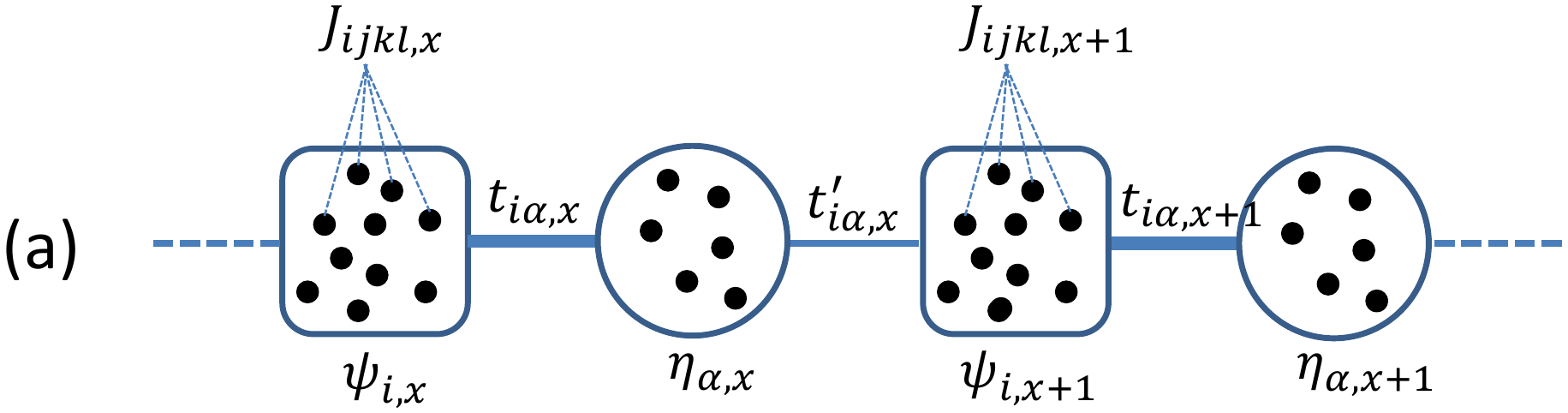}}
\subfigure{\includegraphics[width=6.5cm]{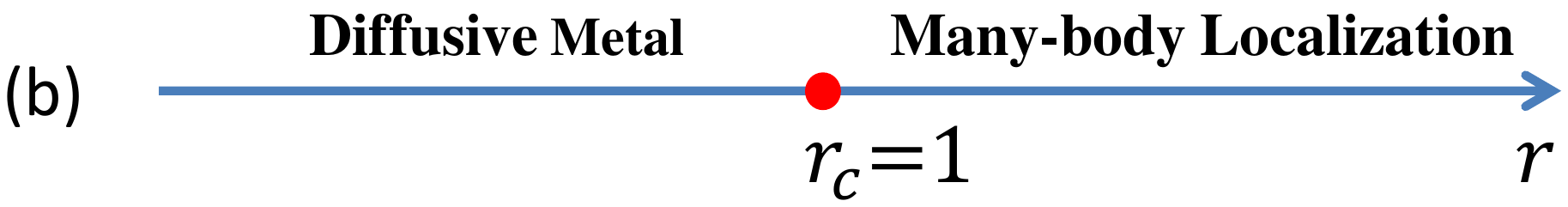}}
\caption{\label{model} ({\bf a}) The 1D generalization of the SYK model consists of $N$ SYK Majorana fermions $\psi_i$ on each site of the $A$ sublattice and $M$ free Majorana fermions $\eta_\alpha$ on each site of the $B$ sublattice. The hopping between two types of fermions is represented by $t_{i\alpha,x}$ and $t'_{i\alpha,x}$. (b) The phase diagram of the 1D model in Eq. (1) as a function of $r$=$M/N$. }
\label{fig1}
\end{figure}

Our generalized SYK model is defined on bipartite lattices. We focus on the case of one-dimensional lattices, while the model can be generalized to any dimensions. As shown in \fig{fig1}(a), each unit cell consists of two sites: one site hosting $N$ Majorana fermions with SYK interactions and the other hosting $M$ free Majorana fermions. Two sublattices are coupled via random hopping. The fermion number ratio is denoted as $r$=$M/N$. Here, we consider the case that both $N$ and $M$ are large but finite while the ratio $r$ is fixed. For $r$$\ll$$1$, SYK physics dominates such that this phase exhibits a finite diffusive constant $D$ and maximal chaos with the Luyapunov exponent satisfying the upper bound $\lambda_L$=$2\pi/\beta$, where $\beta$ is the inverse temperature. It is a diffusive metal, similar to the one studied by Gu {\it et al.} \cite{gu2016}. For $r$$\gg$$1$, the ``free'' Majorana fermions on the 1D lattice dominate over the SYK fermions such that weak SYK interactions are irrelevant around the Anderson-localization ``fixed'' point of free disordered Majorana fermions \cite{footnote2}, leading to MBL.

Consequently, we expect that there should be a dynamic phase transition between a thermal (diffusive) phase and a MBL phase as the ratio $r$ varies from small to large. Indeed, for $r$$<$$1$, our analytical calculations show that the diffusion constant vanishes as $D$$\propto$$(1-r)^{1/2}$ when $r$$\to$1. This implies that a dynamical phase transition to a MBL phase should occur at $r$=$r_c$=$1$. The MBL nature for $r$$>$$r_c$ is further supported by our numerical calculations of the many-body level statistics, which qualitatively changes around $r$=$r_c$: it follows Poisson distribution for $r$$>$$r_c$ but Wigner-Dyson for $r$$<$$r_c$. To the best of our knowledge, it is the first time that a MBL transition is evidenced in a nearly solvable model.

The MBL transition in our generalized SYK models looks qualitatively different from previously studied cases. First, the MBL transition in our generalized SYK model on the 1D lattice occurs between diffusive and MBL phases. This is qualitatively different from previously studied 1D cases, where it was shown by exact diagonalization and real-space renormalization group analysis that a MBL transition can only occur between a subdiffusive phase and a MBL one, both of which have vanishing diffusive constant \cite{lev2015,knap2016,varma2016,altman2015prx,sid2015prx}. Second, because of the local criticality in the generalized SYK models, the critical exponent $\nu$=$0$ at the MBL transition since the spatial correlation length keeps finite at the transition, which seemly violates the Harris criterion $d\nu\!>\!2$ in systems of $d$ spatial dimensions \cite{harris1974, chayes1986, chandran2015}.

{\bf SYK models on 1D lattices:} We first introduce the generalized model on 1D lattices, as shown in \fig{fig1}(a), and consider the cases of more than 1D later. The Hamiltonian of the generalized SYK model in 1D reads:
\bea\label{model}
	H &=& \sum_{x=1}^{L} \Big[ \frac{1}{4!} \sum_{ijkl} J_{ijkl,x} \psi_{i,x} \psi_{j,x} \psi_{k,x}\psi_{l,x} \nn\\
	&&~~~ + \sum_{i\alpha} \big( t_{i\alpha,x} i\psi_{i,x} \eta_{\alpha,x} + t_{i\alpha,x}'  i\eta_{\alpha,x} \psi_{i,x+1} \big) \Big],
\eea
where $\psi_{i,x}$ and $\eta_{\alpha,x}$ are SYK Majorana fermions and free Majorana fermions residing on the $A$ site and the $B$ site of the unit cell $x$, respectively, with $i$=1,$\cdots$,$N$ and $\alpha$=1,$\cdots$,$M$. The number of unit cells in the chain is $L$, and the periodic boundary condition is assumed. The SYK fermions on the $A$ sublattice have on-site all-to-all random four-fermion interactions $J_{ijkl,x}$ with mean zero and variance $\langle J_{ijkl,x}^2 \rangle \!=\! J^2 3!/N^3$. Here, $t$ and $t'$ are nearest neighbor random hopping of Majorana fermions within the same unit cell and between neighboring unit cells, respectively, with mean zero and variance $\langle t_{i\alpha,x}^2 \rangle=t^2/\sqrt{MN}$ and $\langle t_{i\alpha,x}^{\prime2} \rangle= t^{\prime2}/\sqrt{MN}$. Hereafter, we assume $t^\prime$$\ll$$t$. When we take the large-$N$ limit, we keep the ratio $r\equiv \frac{M}{N}$ fixed. Note that the time-reversal symmetry ($\psi$$\rightarrow$$\psi$, $\eta$$\rightarrow$$-\eta$ and $i$$\rightarrow$$-i$) is assumed for the generalized model such that hopping between the same type of fermions is forbidden.

Like in the original SYK model, we use a replica trick to get an effective disorderless model (see the Supplemental Material for details) and introduce bilocal variables: $G^{mm'}_{\psi,x}(\tau_1,\tau_2)$=$\frac{1}{N} \sum_{i=1}^N \psi^m_{i,x}(\tau_1) \psi^{m'}_{i,x}(\tau_2)$ and $G^{mm'}_{\eta,x}(\tau_1,\tau_2)\!=\!\frac{1}{M} \sum_{\alpha=1}^M \eta^m_{\alpha,x}(\tau_1)\eta^{m'}_{\alpha,x}(\tau_2)$, as well as $\Sigma^{mm'}_{\psi,x}(\tau_1,\tau_2)$, $\Sigma^{mm'}_{\eta,x}(\tau_1,\tau_2)$ as Legendre multipliers to implement the above identities, where $m,m'$ are replica indices. At the large-$N$ limit, different replicas do not interact, so the bilocal fields are diagonal in replica indices, i.e., $G^{mm'}=G \delta^{mm'}$ and $\Sigma^{mm'}=\Sigma \delta^{mm'}$. We obtain the following effective action:
\bea\label{action}
	\frac{S}{N} &=&\sum_{x=1}^L\bigg[  -\frac12[\text{tr} \log(\partial_\tau- \Sigma_{\psi,x})+r~\text{tr} \log (\partial_\tau- \Sigma_{\eta,x}) ]  \nn \\
	&& ~~~+ \frac{1}{2} \iint \Big( \Sigma_{\psi,x} G_{\psi,x} + r\Sigma_{\eta,x} G_{\eta,x} - \frac{J^2}{4} G_{\psi,x}^4 \nn\\
	&&~~~~~~~~~~- \sqrt{r}(t^2 G_{\psi,x} G_{\eta,x}+ t^{\prime2} G_{\eta,x} G_{\psi,x+1})  \Big)\bigg],~
\eea
where $G$ and $\Sigma$ are collective bosonic modes and $\iint \equiv \int d\tau_1 d\tau_2$ (integration over two times appears because the replica trick couples fields at different times). The large-$N$ structure is manifest in the effective action above. The saddle-point equations obtained by varying these collective modes are
\bea
	&& G_{\psi,x}^{-1}(i\omega)\!=\!-i\omega \!-\! \Sigma_{\psi,x}(i\omega),~G_{\eta,x}^{-1}(i\omega) \!=\! -i\omega \!-\! \Sigma_{\eta,x}(i\omega), \label{SD1}\\
	&& \Sigma_{\psi,x}(\tau)=J^2 G_{\psi,x}^3(\tau)+ \sqrt{r} [t^2 G_{\eta,x}(\tau)+ t^{\prime2} G_{\eta,x-1}(\tau)],~~~\label{SD2} \\
	&& \Sigma_{\eta,x}(\tau)=[t^2 G_{\psi,x}(\tau)+t^{\prime2} G_{\psi,x+1}(\tau)]/\sqrt{r},\label{SD3}
\eea
where $\tau\!=\!\tau_1\!-\!\tau_2$.
These saddle-point equations are equivalent with Schwinger-Dyson equations obtained from diagrammatic methods \cite{stanford2016a, gu2016}.

{\bf Diffusive metals:} For $r$$\ll$1, it is expected that the SYK fermions dominate over the free Majorana fermions in the infrared \cite{altman2017}. Similar to features of the original SYK model, the time-derivative terms in \Eq{action} or the $-i\omega$ terms in \Eq{SD1} are irrelevant in low energy. Remarkably, Eqs.(\ref{SD1}-\ref{SD3}) in the infrared limit of $\omega\to 0$ are invariant under global (site-independent) reparametrization of time $\tau \rightarrow f(\tau)$,
\bea
 	\tilde{G}_{a,x}(\tau_1,\tau_2) &=& [f'(\tau_1)f'(\tau_2)]^{\Delta_a} G_{a,x}(f(\tau_1),f(\tau_2)),
\eea
where $f'(\tau)$=$\frac{df}{d\tau}$, $a$=$\psi$ or $\eta$, and the scaling dimensions $\Delta_\psi$=$\frac14$, $\Delta_\eta$=$\frac34$. Like in the SYK model, this is an emergent time reparametrization symmetry at low energy that is explicitly broken by high-energy degrees of freedom in the microscopic model [or the time derivative-terms in the effective action \Eq{action}].

Helped by the emergent reparametrization symmetry, we obtain the following solutions of the Schwinger-Dyson equations in the infrared [Eqs.(\ref{SD1}-\ref{SD3}) in the limit of $\omega\to 0$]:
\bea
G^s_{\psi,x}(\tau) &=& \Big( \frac{1-r}{4\pi J^2}\Big)^{1/4} \frac{\sgn(\tau)}{|\tau|^{1/2}},\label{solution1}\\
G^s_{\eta,x}(\tau) &=& \frac{1}{2(t^2+t^{\prime2}) } \Big[ \frac{r^2 J^2}{4\pi^3(1-r)} \Big]^{1/4} \frac{\sgn(\tau)}{|\tau|^{3/2}}.  \label{solution2}
\eea
The solutions above are spatially uniform while nontrivial in the time direction, exhibiting local criticality \cite{gu2016,si2001,faulkner2011}. Note that the saddle-point solutions are valid below a cutoff frequency $\omega_c$ which scales as $\omega_c$$\sim$$(1-r)^{1/6}$ when $r$$\rightarrow$$1$ (see the SM for details).Using the saddle-point solutions, we find that the zero-temperature entropy per unit cell per Majorana fermion is given by (see the SM for details) $\mathcal{S}= \frac{1-r}{1+r} \mathcal{S}_\text{SYK}$,
where $\mathcal{S}_\text{SYK}= \frac{4 \mathcal{C}+\pi \log 2}{8\pi} \approx 0.232$, and $\mathcal{C} \approx 0.916$ is Catalan constant. When $r\rightarrow 1$, the zero temperature entropy vanishes which implies a hint that there is a phase transition at $r_c$=$1$.

Note that the saddle-point solutions of Eqs.(\ref{solution1}-\ref{solution2}) spontaneously break the continuous reparametrization symmetry to SL(2,$R$). Owing to the spontaneous and explicit breaking pattern, site-dependent reparametrization modes $\epsilon_{x}\!=\!f_{x}(\tau)-\tau$ would contribute dominant low-energy fluctuations on top of the saddle-point one, which determine the low-energy physics especially dynamics like transport and the butterfly effect. Note that because $t'\ll t$ the relative reparametrization fluctuation within each unit cell (namely, $f_{\psi,x}$$-$$f_{\eta,x}$) is at high energy and does not affect the physics in the low energy we consider here.

The effective action for the reparametrization modes is given by fluctuations around the saddle-point one, i.e., $S_\text{eff}[f]=S[\tilde{G}(f)]-S[G(\tau)]$, where $\tilde{G}_{a,x}(\tau_1,\tau_2)=f'^{\Delta_a}_{x}(\tau_1)f'^{\Delta_a}_{x}(\tau_2) G_{a}(f_{x}(\tau_1),f_{x}(\tau_2))$ is the Green's function of $a$=$\psi,\eta$ fermions associated with the spatially dependent time reparametrization $f_{x}(\tau)$=$\tau+\epsilon_{x}$. Note that though the saddle-point solution of $G_{a,x}$ in Eqs.(\ref{solution1}-\ref{solution2}) is homogenous, its fluctuation associated with the reparametrization modes is generically inhomogeneous. By assuming weak reparametrization $\epsilon_x$ as well as performing $\varepsilon$ expansion and series summation (see the Supplemental Material for details), we obtain the effective action up to the quadratic in $\epsilon$,
\bea
\frac{S_\text{eff}}{N} &=& \frac{\pi}{\beta} \sum_{n,p}\Big( \alpha_1 |\omega_n| + \alpha_2 p^2 \Big) |\omega_n| \Big[ \omega_n^2- \big(\frac{2\pi}{\beta}\big)^2 \Big] |\epsilon_{\omega_n,p}|^2,~~~
\eea
where $\omega_n$=$2\pi n/\beta$ is the Matsubara frequency, $p$ is momentum, and $\epsilon_x(\tau)$=$\frac{1}{\sqrt{L}\beta} \sum_{n,p} \epsilon_{\omega_n,p} e^{-i\omega_n\tau+ipx}$. As shown in the Supplemental Material, $\alpha_1$=$\frac{1}{64\pi^2} \Big(\frac{\sqrt{1-r}}{J}$+$ \frac{J}{t^2+t^{\prime2}} \sqrt{\frac{r^3}{1-r}}\Big)$ and $\alpha_2$=$\frac{1}{128\pi } \frac{r t^{\prime2}}{t^2+t^{\prime2}}$. Since $J$ and $t$ are both relevant at the UV Gaussian point, they increase as energy scales lower. Thus, $\alpha_1$ becomes extremely small due to the emergent reparametrization symmetry, while $\alpha_2$ is also small in the homogenous limit, i.e., $t'\ll t$. These lead to strong fluctuations of reparametrization modes which dominate the low-energy dynamics.

Having obtained the effective action for the reparametrization modes, we are ready to calculate their contributions to energy transport and OTOC in the limit of $N$$\gg$$\beta J$$\gg$1. The energy density for small momentum is given by $T_{\omega_n,p}$=$\frac{iN\alpha_1}{4\pi} \omega_n[\omega_n^2- \big(\frac{2\pi}{\beta}\big)^2]\epsilon_{\omega_n,p}$. Using the effective action for reparametrization modes, the real-frequency correlator  (see the Supplemental Material for details) $\label{energycorr}\langle T_{-\omega,-p} T_{\omega,p} \rangle = \frac{N\alpha_1}{8\pi \beta^2} \frac{Dp^2}{-i\omega+D p^2}$, where the diffusive constant $D$ is
\bea\label{diff-const}
D= \frac{\pi}{2} \frac{r\sqrt{1-r} J t^{\prime2}}{(1-r) (t^2+t^{\prime2})+ r^{\frac32} J^2}.
\eea
Some remarks come with this expression for diffusive transport of energy. First, when $t^\prime=0$, different unit cells decouple from each other, and the diffusive constant $D$ vanishes as expected. On the other hand, when $r$$\rightarrow$$0$, the free Majorana fermions vanish, and the system becomes decoupled islands of SYK Majorana fermions and cannot conduct energy. A more interesting observation is that when $r$$\rightarrow$$1$, the diffusive constant scales as $D \propto (1-r)^{1/2}$, and we expect the system enters a localized phase.

We are now in a position to calculate the OTOC. Consider the following four-point correlation function
\bea
&&F_{\psi\psi,xx'}(\tau_1\tau_2\tau_3\tau_4)\!=\! \frac1{N^2}\!\langle T_\tau\! \sum_{ij}\psi_{i,x}(\!\tau_1\!) \psi_{i,x}(\!\tau_2\!) \psi_{j,x'}(\!\tau_3\!) \psi_{j,x'}(\!\tau_4\!)\rangle \nn \\
&&~~~~~~~~~~~~~~~~~~= G^s_\psi(\tau_1\tau_2)G^s_\psi(\tau_3\tau_4)+ \frac{1}{N} \mathcal{F}_{\psi\psi,xx'}(\tau_1\tau_2\tau_3\tau_4), \nn \\
\eea
where $T_\tau$ denotes imaginary time ordering, $G^s_\psi$ is given by the saddle-point solutions in Eqs.(\ref{solution1}-\ref{solution2}), and $\mathcal{F}_{\psi\psi,xx'}$ is the connected part coming from the fluctuations around the saddle-point, $\mathcal{F}_{\psi\psi,xx'}(\tau_1\tau_2\tau_3\tau_4) \equiv \langle \delta G_{\psi,x}(\tau_1\tau_2) \delta G_{\psi,x'}(\tau_3\tau_4) \rangle$ and is dominated by the reparametrization modes. Similar calculations apply to the  OTOC of other operators. In order to evaluate the OTOC, let $\tau_1$=$\beta$+$it$, $\tau_4$=$\frac{3\beta}{4}$, $\tau_2$=$\frac{\beta}{2}$+$it$, $\tau_3$=$\frac{\beta}{4}$, and we arrive at (see the Supplemental Material for details)
\bea
	\frac{\mathcal{F}_{ab,xy}(\tau_1\tau_2\tau_3\tau_4)}{G^s_a(\frac{\beta}2) G^s_b(-\frac{\beta}2)} &\propto& -\frac{\Delta_a\Delta_b}{4\pi\sqrt{\alpha_1 \alpha_2}} \sqrt{\frac{\beta}{2\pi}} e^{\frac{2\pi}{\beta} \big( t-\frac{|x-y|}{v_B} \big) },
\eea
with $v_B^2$=$\frac{2\pi}{\beta} D$ and $a,b$=$\psi,\eta$. We first note that the quantum analog of the Lyapunov exponent defined by OTOC in this phase still saturates the bound $\lambda_L$=$\frac{2\pi}{\beta}$ \cite{stanford2016b}. Second, the butterfly velocity, Lyapunov exponent, and diffusive constant here satisfy a simple and elegant relation: $D$=$\frac{v_B^2}{\lambda_L}$ \cite{hartnoll,blake2016prl}. Such a relation was previously obtained in incoherent black holes \cite{blake2016, blake2017} and higher-dimensional generalizations of the SYK model \cite{gu2016,sachdev2016,footnote}. As the butterfly velocity $v_B$$\propto$$(1-r)^\frac14$ is vanishing for $r$$\to$1, it further indicates that the system shall undergo a localization transition as $r$ crosses the critical value $r_c$=1.

{\bf MBL phase:}
For $r \gg 1$, it is expected that the Anderson localization of ``free'' Majorana fermions for large but finite $N$ dominate in determining low-energy physics and the SYK interaction $J$ is irrelevant. Consequently, the system should fall into a localized phase \cite{footnote2}. Similar to the case of $r\ll 1$, we also make a translational invariant ansatz for $r\gg 1$, with which the saddle-point equation can be approximated by
\bea
	&& G_{\psi}^{-1}=-i\omega - \Sigma_{\psi}, ~~~~~~G_{\eta}^{-1}= -i\omega- \Sigma_{\eta}, \\
	&& \Sigma_{\psi}= \sqrt{r} \tilde t^2 G_{\eta},~~~~~~~~~~ \Sigma_{\eta}= \tilde t^2 G_{\psi}/\sqrt{r},~~~
\eea
where ${\tilde t}^2\equiv t^2+t'^2$. The exact solutions of the above Schwinger-Dyson equations are obtained in the Supplemental Material. Here, let's explicitly expand the inverse propagators $G_a^{-1}$ around small frequency:
\bea
	G_\eta^{-1}&\!=\!&-\frac{r}{r-1} i\omega-\frac{r^{3/2}}{(r-1)^3 \tilde t^2} (i\omega)^3+ O(\omega^5),\label{Gweta}\\
	G_\psi^{-1}&\!=\!& \frac{r-1}{\sqrt{r}} \frac{1}{i\omega}\!-\! \frac{r}{r-1} i\omega \!-\! \frac{r^{3/2}}{(r-1)^3 \tilde t^2} (i\omega)^3 \!+\! O(\omega^5). ~~~\label{Gwpsi}
\eea
Although in $G^{-1}_\eta$ the bare term $\propto -i\omega$ is renormalized by a factor $\frac{r}{r-1}$, its self-energy is subdominant at low energy, indicating the free Gaussian fixed point of $\eta$ Majorana fermions is stable. (In the limit of $r$$\rightarrow$$\infty$, $G_\eta^{-1}$$\rightarrow$$-i\omega$, as expected from a free theory.) However, for the $\psi$ fermions, the self-energy actually dominates the behavior of $G_\psi$ in low energy, which generates a large anomalous dimension to $\psi$. For simplicity, we keep the leading term in \Eq{Gwpsi} and make a Fourier transformation,
\bea
	G_\psi(\tau)= \frac{\sqrt{r}}{2(1-\gamma)(r-1)}\frac{\text{sgn}(\tau)}{|\tau|^2},
\eea
where $\gamma \approx 0.577$ is the Euler-Gamma constant. From the propagator of $\psi$ fermions, one deduces its scaling dimension $[\psi]$=$1$, as expected from the $r$$\gg$1 free fixed point.

Now we explore effects of the SYK interaction $J$. Including $J$ terms leads to a correction to the self-energy, $\delta \Sigma_\psi(\tau)$=$J^2 G_\psi(\tau)^3$=$ \frac{r^{3/2}J^2}{8(1-\gamma)^3(r-1)^3}\frac{\text{sgn}(\tau)}{|\tau|^6}$.
By a Fourier transformation, $\delta \Sigma_\psi(i\omega)\propto\omega^5$, which is subdominant in low energy, compared with leading terms in \Eq{Gwpsi}. The same is true for $\delta \Sigma_\eta(i\omega)$. Thus, we can conclude the free fixed point with $[\eta]$=$0$ and $[\psi]$=$1$ is stable against weak interaction $J$, which self-justifies the assumption we have made. One important consequence is that, as all levels of the free Majorana fermions with random hopping are localized for large and finite $N$ \cite{footnote2,herbert1971}, MBL emerges in the presence of the weak but irrelevant SYK interaction $J$ \cite{altman2013}. It is consistent with vanishing diffusive constant for $r$$>$$1$. Note that the system has $(M-N)L$ single-particle zero modes localized on the B sites for $r>1$. However, these localized states do not change the MBL phase because they are isolated from the rest of the many-body states and only cause macroscopic degeneracies. One way of removing these extensive localized zero modes but preserving the low-energy physics is to add weak quadratic couplings on B sites, as we show in the Supplemental Material.

\begin{figure}[t]
\subfigure[]{\label{distr1}\includegraphics[width=2.81cm]{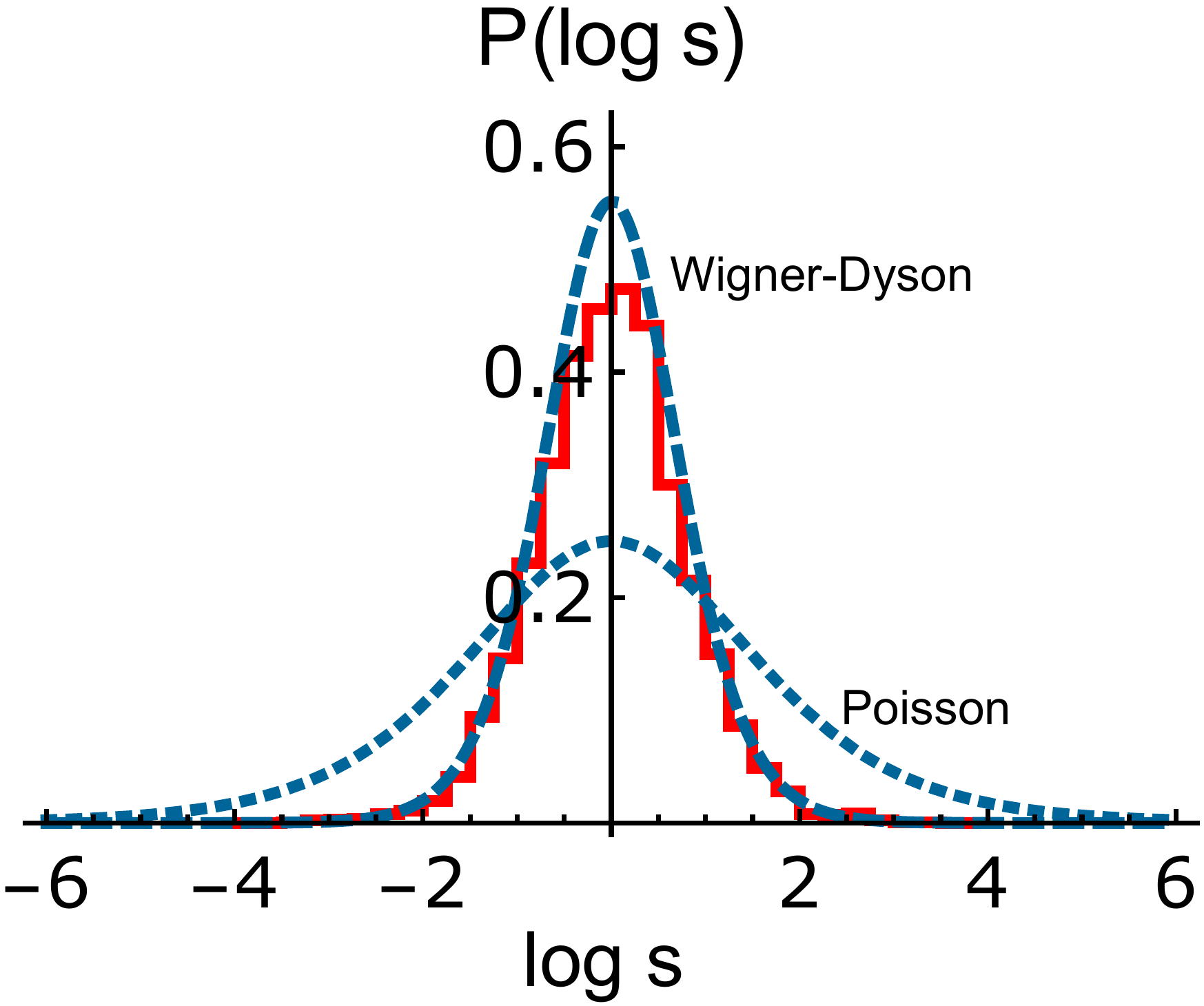}}
\subfigure[]{\label{distr2}\includegraphics[width=2.81cm]{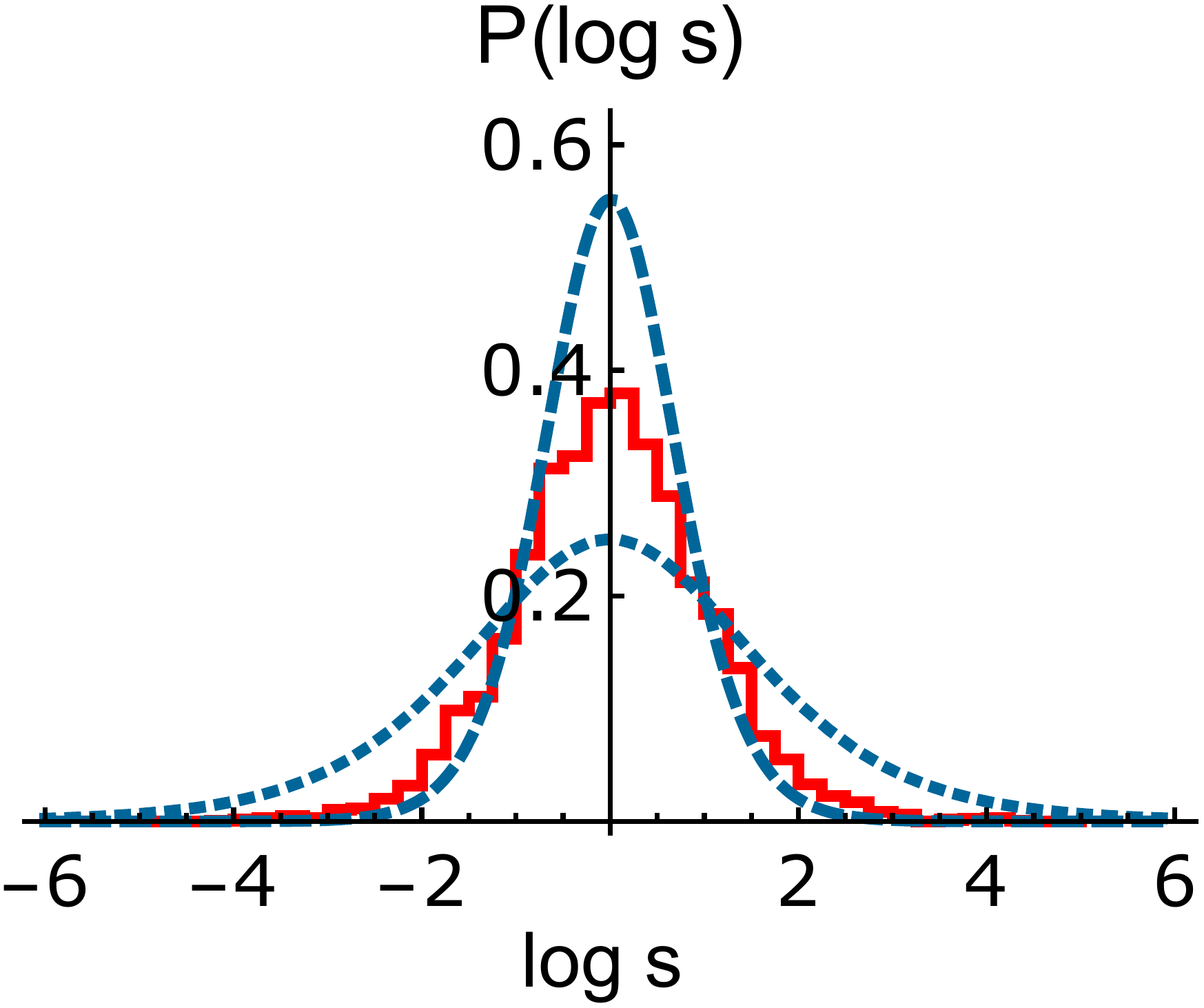}}
\subfigure[]{\label{distr3}\includegraphics[width=2.81cm]{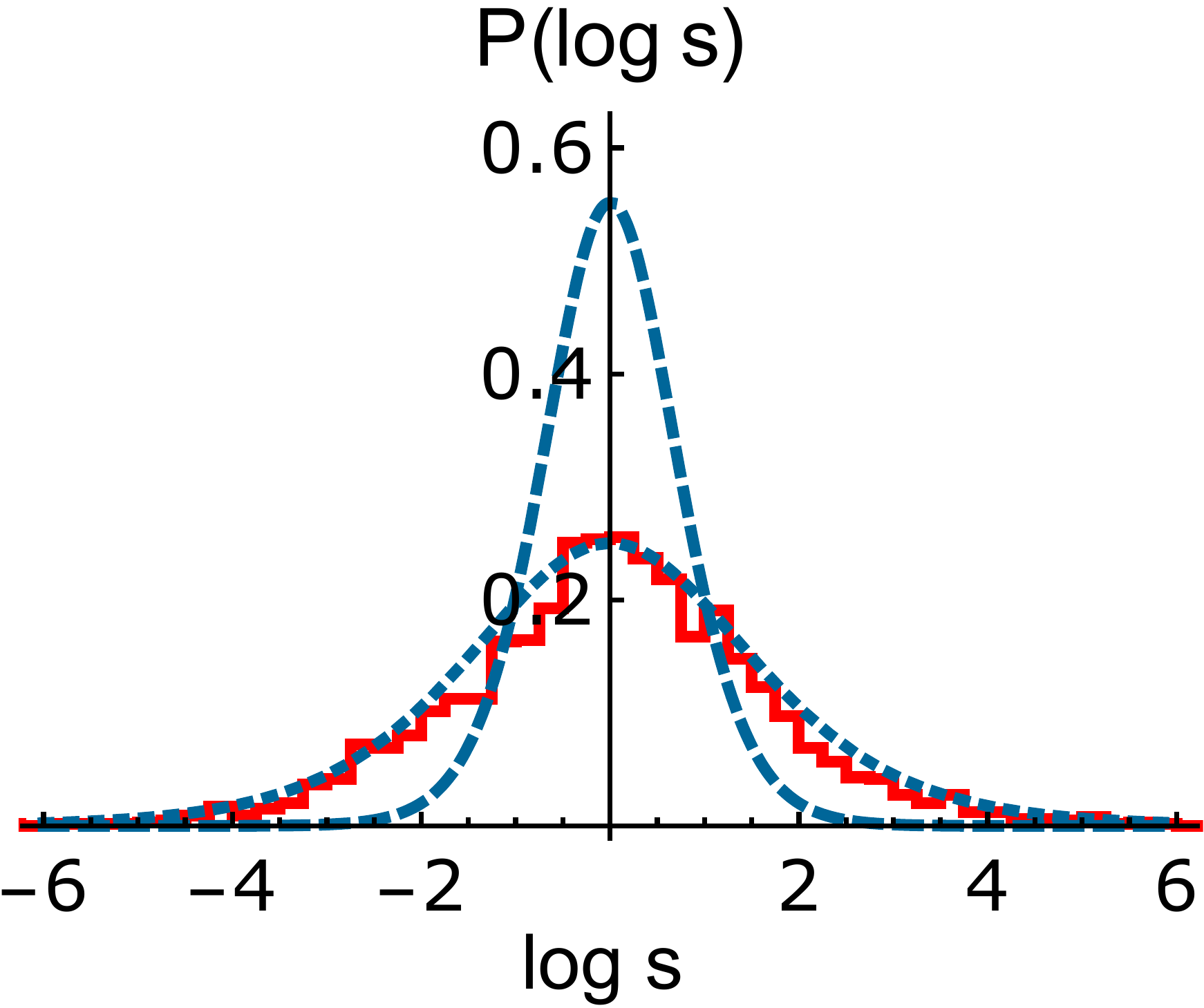}}
\caption{\label{distribution} The distribution of level-spacing ratios for the cases of $(N$,$M)$=(6,4), (5,5) and (4,6) are shown in (a), (b) and (c), respectively.  The results (red solid line) are obtained by exactly diagonalizing the generalized SYK model on the six-site chain with $N$+$M$=10 Majorana fermions in each unit cell and with $J$=$t$=1, $t'$=0.5. The Wigner-Dyson distribution (dashed line) implies thermalization while Poisson distribution (dotted line) implies MBL. }\label{dist}
\end{figure}

{\bf Numerical evidences of MBL transitions:} We now show numerical evidence of such a phase transition between the thermal and MBL phases. For a MBL phase, its level statistics satisfies the Poisson distribution according to the Berry-Tabor conjecture \cite{berry1977} while a thermal phase's level statistics follows the Wigner-Dyson (WD) distribution. Suppose $\{E_n\}$ denotes many-body eigenstate energies in an ascending order and the level spacings between adjacent eigenstates are $\Delta_n$=$ E_{n+1}$$-$$E_n$ with $\Delta_n$$ \ge $$0$. The ratio between two consecutive gaps $s_n$=$\frac{ \Delta_{n+1}}{\Delta_{n}}$ can be employed to characterize the level statistics \cite{huse2007,atas2013}. The distribution of ratios in MBL phases follows Poisson level statistics $p(s)$=$\frac{1}{(1+s)^2}$, while in thermalized phases, it follows WD level statistics $p(s)$=$\frac{81\sqrt{3}}{4\pi} \frac{(s+s^2)^2}{(1+s+s^2)^{4}}$ (assuming Gaussian unitary ensemble).

Following Ref. \cite{you2016}, we plot the distribution of $\log s$, i.e., $P(\log s)= p(s)s$, as shown in Fig. \ref{distribution}. The data are obtained from exactly diagonalizing the model with $J$=$t$=1, $t'$=0.5 on a six-site chain with $N$+$M$=10 Majorana fermions per unit cell. The distribution for $(N,M)$=(6,4), (5,5), (4,6) is shown in \fig{dist}(a,b,c), respectively. When $N$$>$$M$ (namely, $r$$<$$1$), the distribution in \fig{distr1} follows that of WD; when $N$$<$$M$ (namely, $r$$>$$1$), the distribution in \fig{distr3} follows that of Poisson. When $N$=$M$ (namely, $r$=$r_c$=$1$), the distribution in \fig{distr2} is in transition between Poisson and WD. Our numerical results imply that a dynamic transition from a thermal to a MBL phase occurs around $r$=$1$. As mentioned before, for $r>1$, each many-body energy level has an extra degeneracy due to the presence of the single-particle zero modes localized on B sites. In the calculation of energy level statistics, we have ignored these trivial degeneracy. The degeneracy can be lifted by adding weak quadratic couplings on B sites, and for such modified case we also calculated the level statistics and obtained the qualitatively same results, as shown in the Supplemental Material.

\begin{figure}[t]
	\subfigure[]{\includegraphics[height=3.8cm]{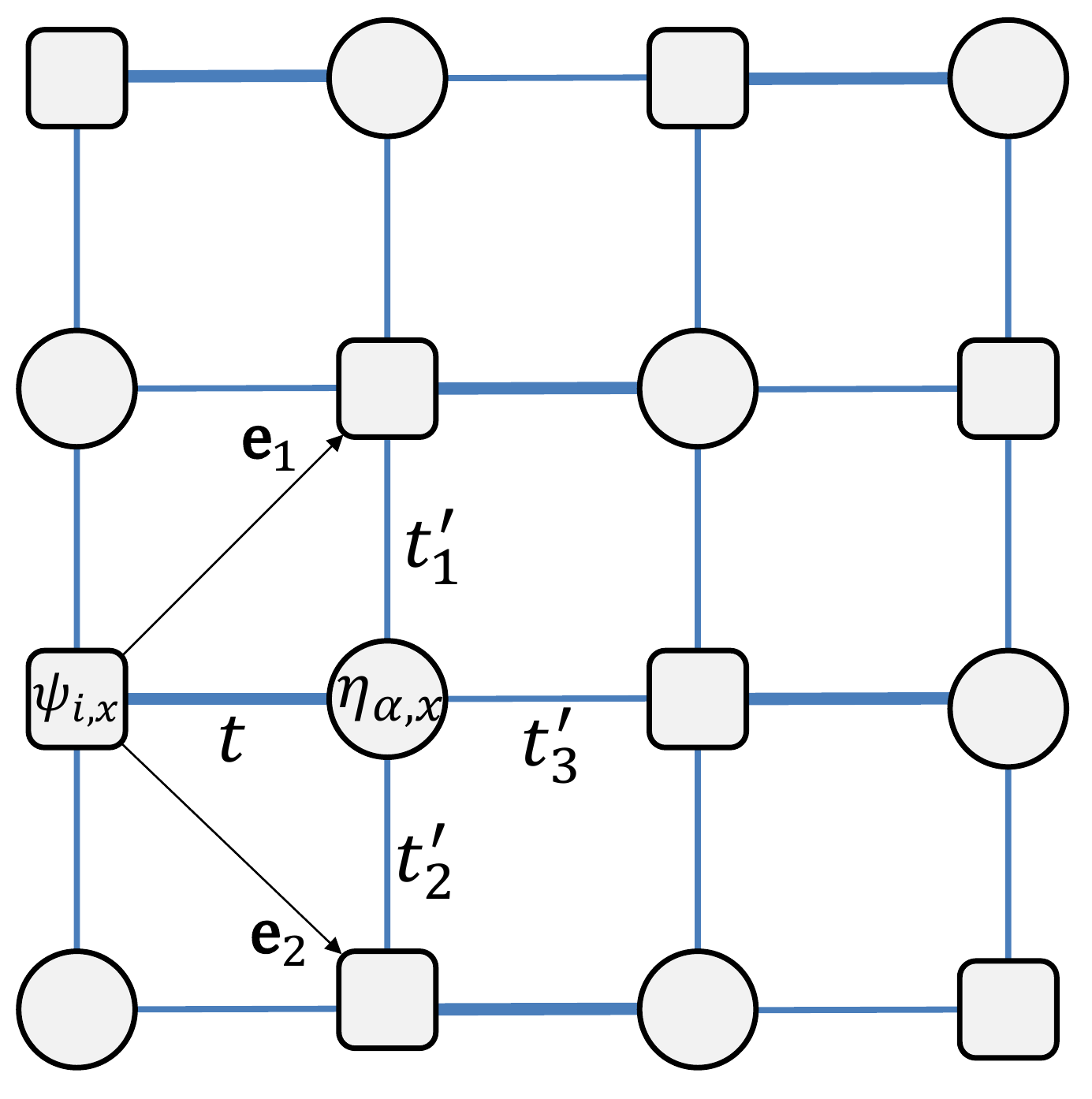}}~~~~~~
	\subfigure[]{\includegraphics[height=3.8cm]{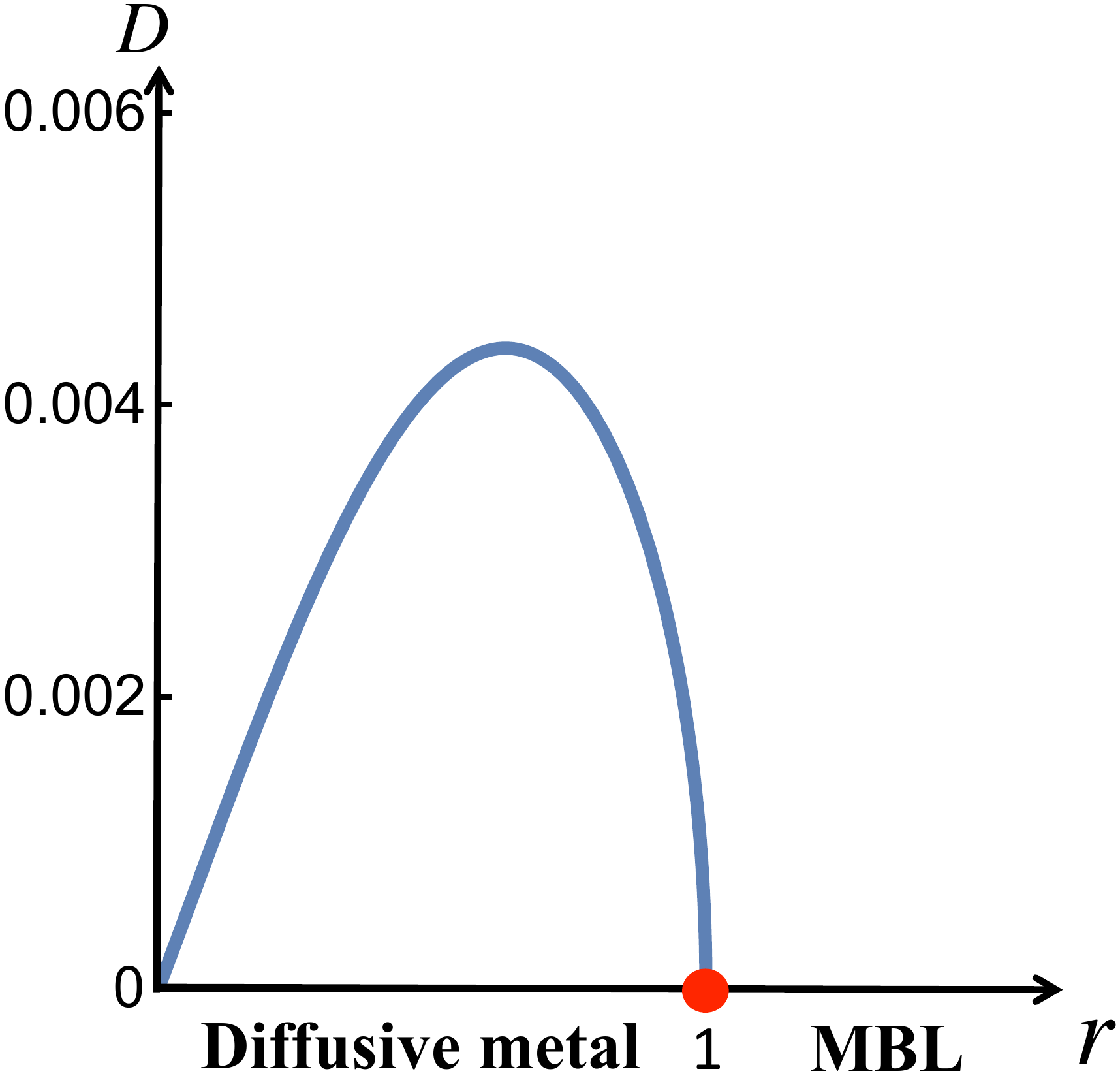}}
	\caption{\label{2d}(a) The generalized SYK model on the square lattice. Each unit cell consists of two sites represented by a square and a disk, where $N$ SYK Majorana fermions and $M$ free Majorana fermions reside, respectively. $t$ denotes the variance of random hopping within a unit cell, while $t'$ denotes that between neighboring unit cells. (b) The energy diffusive constant $D$ along the ${\bf e}_1$ or ${\bf e}_2$ direction as a function of $r$. We use the parameter $J$=$t$=1, $t_1'$=$t_2'$=0.1, $t_3'$=0. }
\end{figure}

{\bf SYK model on 2D lattices:} Our construction of the SYK models in 1D can be straightforwardly generalized to more than 1D. For instance, we consider the generalization to the square lattice as shown in \fig{2d}. Each unit cell consists of two sites represented by a square and a disk, where $N$ SYK Majorana fermions and $M$ free Majorana fermions reside, respectively. The model is given by
\bea
	H &=& \sum_{\bf x} \Big[ \frac{1}{4!} \sum_{ijkl} J_{ijkl,{\bf x}} \psi_{i,{\bf x}} \psi_{j,\bf x} \psi_{k,\bf x}\psi_{l,\bf x} \nn\\
	&& + \sum_{i\alpha} \Big( t_{i\alpha,\bf x} i\psi_{i,\bf x} \eta_{\alpha,\bf x} + \sum_a t_{i\alpha,{\bf x}a}'  i\eta_{\alpha,\bf x} \psi_{i,{\bf x}+{\bf b}_a} \Big) \Big], \nn
\eea
where ${\bf x}$ represents unit cells, and ${\bf b}_a$ label the vectors connecting neighboring unit cells with ${\bf b}_1$=${\bf e}_1$=$(1,0)$, ${\bf b}_2$=${\bf e}_2$=$(0,1)$, and ${\bf b}_3$=$(1,1)$. Similarly, $\langle J_{ijkl,{\bf x}}^2\rangle$=$3!J^2/N^3$, $\langle t_{i\alpha,{\bf x}}^2 \rangle$=$t^2/\sqrt{MN}$, and $\langle t_{i\alpha,{\bf x}a}'^2 \rangle$ =$ t^{\prime2}_a/\sqrt{MN}$. (Note that the limit of $t'_3$=$0$ corresponds to the honeycomb lattice). The analysis of the generalized model on 2D and higher-dimensional lattices goes like the 1D chain case. For $r$$<$$1$, the generalized models on 2D lattices possess similar features, including diffusive energy transport, zero-temperature entropy, and maximum quantum chaos, which are the same as the model on a 1D chain. For instance, the diffusive constant in 2D as a function of $r$ is also given by \Eq{diff-const}, which is plotted in \fig{2d}(b). For $r$$\to$$0$, $D\propto r$ because in diffusive metal, the SYK Majorana fermions diffuse via free Majorana fermions; while for $r$$\rightarrow$$1$, $D\propto (1-r)^{1/2}$, which indicates that the system could undergo a dynamical transition into a MBL phase.

{\bf Concluding remarks:} We have shown that the MBL transition in the generalized SYK models is qualitatively distinct from previously studied ones in other models like the XXZ model in various ways. Intuitively, we think that the qualitative differences are mainly due to the large-$N$ degrees of freedom on each site in the generalized SYK models. In the large-$N$ limit, due to the all-to-all interactions, we can define an effective dimensions $d_\text{SYK}$$\to$$\infty$ such that the effective dimensions of the generalized model on the $d$-dimensional lattice is $d^\ast$=$d_\text{SYK}$$+$$d$, which approaches infinity. As a consequence, for the SYK model on the $d$=$1$ lattice, there is no subdiffusive phase around the MBL transition because its effective space dimension $d^\ast$ is much larger than 1. Moreover, the Harris criterion is not violated by $\nu$=$0$ when $d^\ast$ is considered as the effective space dimension.

Note that there are questions that remain open. To inspire readers, we provide a few here. First, what is the critical theory governing this MBL transition? Our analysis cannot be applied directly at $r$=$1$, and the critical theory remains unknown. Second, is time-reversal symmetry spontaneously broken in the MBL phase ($r$$\gtrsim$$1$)? Although we have shown that the $J$ term is irrelevant when $r$$\gg$$1$, it is possible to be dangerously irrelevant for $r$$\gtrsim$$1$. Third, how robust is the critical point when other types of interactions are included in the model?

{\it Acknowledgements}: We would like to thank Xin Dai, Yingfei Gu, David Huse, Xiaoliang Qi, Cenke Xu, and Shixin Zhang for helpful discussions. This work is supported in part by the MOST of China under Grant No. 2016YFA0301001 (H. Y.) and by the NSFC under Grant No. 11474175 (S.-K. J. and H. Y.).

{\it Note added}: After the completion of the present work, we became aware of an upcoming work \cite{cenke} studying a different generalization of the SYK model with a zero-temperature insulating phase (but not MBL).

\begin{widetext}
\section{Supplemental Material}
\renewcommand{\theequation}{S\arabic{equation}}
\setcounter{equation}{0}
\renewcommand{\thefigure}{S\arabic{figure}}
\setcounter{figure}{0}
\renewcommand{\thetable}{S\arabic{table}}
\setcounter{table}{0}

\subsection{A. Replica action}
The Hamiltonian in main text is given by
\bea
	H= \sum_{x=1}^{L} \Big[ \frac{1}{4!} \sum_{ijkl} J_{ijkl,x} \psi_{i,x} \psi_{j,x} \psi_{k,x}\psi_{l,x}+ \sum_{i\alpha} \Big( t_{i\alpha,x} i\psi_{i,x} \eta_{\alpha,x} + t_{i\alpha,x}'  i\eta_{\alpha,x} \psi_{i,x+1} \Big) \Big],
\eea
where $\psi$ and $\eta$ are SYK Majorana fermions and free Majorana fermions, $J_{ijkl,x}$, $t_{i\alpha,x}$ and $t_{i\alpha,x}'$ are independent random couplings with zero mean and variance $\langle J_{ijkl,x}^2 \rangle= J^2 3!/N^3$, $\langle t_{i\alpha}^2 \rangle=t^2/\sqrt{MN} $, and $\langle t_{i\alpha}^{\prime2} \rangle= t^{\prime2}/\sqrt{MN}$. Note that $N$, $M$ are the numbers of SYK Majorana fermions and free Majorana fermions in each sub-lattice respectively, while $L$ is the number of unit cells in the chain. Replica trick utilizes the identity $\log Z= \lim_{n\rightarrow 0} \frac{e^{n\log Z}-1}{n}= \lim_{n\rightarrow 0} \frac{Z^n-1}{n}$. Instead of disorder averaging the logarithm of partition function which is difficult to do, one averages over $n$ copies of the system, then take $n \rightarrow 0$ limit. After the disorder average, the replica action is
\bea
	S&=& \sum_{m,x} \int (\frac12 \psi_{i,x}^m \partial_\tau \psi_{i,x}^m +  \frac12 \eta_{\alpha,x}^m \partial_\tau \psi_{\alpha,x}^m  ) \nn\\
	  && +\sum_{m,m',x} \iint \Big[ - \frac{J^2}{8N^3} \Big(\psi_{i,x}^m \psi_{i,x}^{m'} \Big)^4- \Big( \frac{t^2}{2\sqrt{MN}}\psi_{i,x}^m \psi_{i,x}^{m'} \eta_{\alpha,x}^m \eta_{\alpha,x}^{m'}+\frac{t^{\prime2}}{2\sqrt{MN}} \psi_{i,x+1}^m \psi_{i,x+1}^{m'} \eta_{\alpha,x}^m \eta_{\alpha,x}^{m'} \Big) \Big],
\eea
where $m, m'$ is replica index. As explained in main text, here we consider the diagonal parts. Introduce two collective modes, $G_{\psi,x}(\tau_1,\tau_2) =\frac{1}{N} \sum_i \psi_{i,x} (\tau_1) \psi_{i,x} (\tau_2)$, $G_{\eta,x}(\tau_1,\tau_2) =\frac{1}{M} \sum_\alpha \eta_{\alpha,x} (\tau_1) \eta_{\alpha,x} (\tau_2)$, and corresponding Legendre multipliers $\Sigma_\psi$, $\Sigma_\eta$, one arrives at
\bea
	\frac{S}{N} &=& \frac{1}{2} \sum_x  [ -\text{tr} \log(\partial_\tau- \Sigma_{\psi,x}) - r\text{tr} \log (\partial_\tau- \Sigma_{\eta,x}) ]  \nn \\
	&& + \frac{1}{2} \int d\tau_1 d\tau_2 \sum_x \Big[ \Sigma_{\psi,x} G_{\psi,x} + r\Sigma_{\eta,x} G_{\eta,x} - \frac{J^2}{4} G_{\psi,x}^4- \sqrt{r}t^2 G_{\psi,x} G_{\eta,x}-\sqrt{r}t^{\prime2} G_{\eta,x} G_{\psi,x+1}  \Big],
\eea
where Majorana fermions are integrated out.

\subsection{B. Two-point functions in diffusive metal}

The solutions of saddle point function in frequency domain are given by Fourier transforming the time domain solutions, i.e.,
\bea
	G_\psi(i \omega)=i [(1-r)\pi]^{1/4} \frac{\sgn(\omega)}{\sqrt{J|\omega|}}, ~~~~~ G_\eta(i\omega)=\frac{i\sqrt{r}\text{sgn}(\omega)}{(t^2+t^{\prime2})[(1-r)\pi]^{1/4}}  \sqrt{J|\omega|}.
\eea
While finite temperature solutions are obtained by substitution $\tau \rightarrow \tan \frac{\pi \tau}{\beta}$, i.e., $ G_a(\tau)=\Lambda_a \frac{\text{sgn}(\tau)}{|\frac{\beta}{\pi}\sin\frac{\pi\tau}{\beta}|^{\Delta_a}}$. Here $\Lambda_\psi \equiv \Big( \frac{1-r}{4\pi J^2}\Big)^{1/4}$, $\Lambda_\eta=\frac{1}{2(t^2+t^{\prime2})} \Big( \frac{r^2 J^2}{4\pi^3(1-r)} \Big)^{1/4}$, $\Delta_\psi=\frac14$ and $\Delta_\eta=\frac34$. Following the same lines in \cite{altman2017}, the cutoff frequency for validity of these conformal solutions are given by $\omega_c = \text{min}(\omega_1,\omega_2)$, where
\bea	
	\omega_1 \approx \frac{J}{2\sqrt{\pi}}(1-r)^{3/2}+ \frac{J}{2\sqrt{\pi}} \frac{r^2}{(1-r)^{1/2}}, ~~~~~~\omega_2 \approx \Big(\frac{\sqrt{\pi}}{2} \frac{(t^2+t^{\prime2})^2}{J} \frac{\sqrt{1-r}}{r} \Big)^{1/3}.
\eea
When $r$ approach the transition point $r_c=1$, $\omega_c \sim (1-r)^{1/6} \rightarrow 0$, the saddle point solutions break down.

\subsection{C. Zero temperature entropy}
In this subsection, we use $I$ to denote action while $S$ denotes entropy to avoid confusion. The homogenous saddle-point action generalized to $q$-body interaction is given by
\bea
	\frac{I}{NL} &=& \frac{1}{2} [ -\text{tr} \log(- \Sigma_{\psi}) - r\text{tr} \log (- \Sigma_{\eta}) ] + \frac{1}{2} \int d\tau_1 d\tau_2 \Big[ \Sigma_{\psi} G_{\psi} + r\Sigma_{\eta} G_{\eta} - \frac{J^2}{q} G_{\psi}^q - \sqrt{r}\tilde{t}^2 G_{\psi} G_{\eta} \Big]. \label{action_I}
\eea
where time derivative terms are neglected since they are irrelevant in zero temperature. Equations of motion are
\bea
	G_\psi \ast \Sigma_\psi=-1,~~ G_\eta \ast \Sigma_\eta=-1,~~ \Sigma_\psi=J^2 G_\psi^{q-1}+ \sqrt{r} \tilde{t}^2 G_\eta,~~ \Sigma_\eta= \frac{1}{\sqrt{r}} \tilde{t}^2 G_\psi,
\eea
from which we have $J^2 G_\psi\ast G_\psi^{q-1}=-(1-r)$ with the solutions
\bea
	G_\psi(\tau)=b \frac{\text{sgn}(\tau)}{|\tau|^{2\Delta}}, ~~G_\eta= \frac{\sqrt{r}J^2}{(1-r)\tilde{t}^2} G_\psi^{q-1}, ~~ \Sigma_\psi= \frac{J^2}{1-r} G_\psi^{q-1}, ~~ \Sigma_\eta= \frac{\tilde{t}^2}{\sqrt{r}} G_\psi
\eea
where $J^2 b^q \pi= (1-r)(\frac12 -\Delta) \tan \pi \Delta$, and $\Delta=\frac1q$. Free energy is given by $F=TI$, from which we can get entropy through \bea
S=-\frac{\partial F}{\partial T}=- I- T\sum_\alpha \Big(\frac{\delta I}{\delta G_\alpha} \frac{\partial G_\alpha}{\partial T}+ \frac{\delta I}{\delta \Sigma_\alpha} \frac{\partial G_\alpha}{\partial T} +\frac{\partial I}{\partial T} \Big), \label{entropy}
\eea
The second term of Eq. (\ref{entropy}) vanishes when one plugs the solutions; the zero temperature entropy is given by $S_0=- \lim_{T\rightarrow 0} I$. Plug the solutions into the second term of the action Eq. (\ref{action_I}), we have
\bea
	\frac12 J^2 \Big[ \frac{1}{1-r}- \frac{1}{q^2} \Big] \int d\tau_1 d\tau_2 G^q_\psi(\tau_1,\tau_2) = \frac12 \beta b^q J^2 \Big[ \frac{1}{1-r}- \frac{1}{q^2} \Big] \int_0^\beta d\tau \frac{1}{\big( \frac{\beta}{\pi} \sin \frac{\pi \tau}{\beta}\big)^2}=0.
\eea
The vanishing of above integral can be seen by analytical continuing $\tau \rightarrow \tau+i t$. Thus only the first term in saddle point action is not vanishing,
\bea
	S= \frac{NL}{2} [\text{tr} \log(- \Sigma_{\psi}) + r\text{tr} \log (- \Sigma_{\eta}) ]= \frac{NL}{2} [\sum_n \log(- \Sigma_{\psi}(i\omega_n))+ r\sum_n \log (-\Sigma_{\eta}(i\omega_n)) ].
\eea
Fourier transforming the self-energy leads to
\bea
	\Sigma_\psi(w)= C_\psi i \text{sgn}(\omega) |\omega|^{1-2\Delta}, ~~~\Sigma_\eta(w)= C_\eta i \text{sgn}(\omega) |\omega|^{2\Delta-1}
\eea
where $C_\psi$ and $C_\eta$ are two constants independent of $\Delta$. These constants are not important since we will take derivative with respect to $\Delta$. According to Ref. \cite{stanford2016a}, the zero temperature entropy for SYK model satisfies
\bea
	\frac{\partial \mathcal{S}_\text{SYK}(\Delta)}{\partial \Delta}=-\pi (\frac12-\Delta)\tan \pi \Delta.
\eea
where $\mathcal{S}_\text{SYK}= \frac{S_\text{SYK}}{N}$. Here, analogous to SYK model, we have
\bea
	&& \frac{1}{NL}\frac{\partial S_{\psi}(\Delta)}{\partial \Delta}=-\pi (\frac12-\Delta)\tan \pi \Delta, ~~ \frac{1}{NL}\frac{\partial S_{\eta}(\Delta)}{\partial \Delta}=-r\pi (\frac12-\Delta)\tan \pi (1-\Delta), ~~\frac{\partial \mathcal{S}(\Delta)}{\partial \Delta}= - \frac{1-r}{1+r} \pi (\frac12-\Delta)\tan \pi \Delta \nn\\
\eea
and
where $\mathcal{S}=\frac{1}{(N+M)L} (S_{\psi}+S_{\eta})$. The boundary condition can be set by $\mathcal{S}(\frac12)=0$ because when $q=2$, the system has only quadratic term and unique ground state. Then for our interest case, $\Delta=\frac14$, the ground state entropy is given by
\bea
	\mathcal{S}= \frac{1-r}{1+r} \mathcal{S}_\text{SYK}
\eea
where $\mathcal{S}_\text{SYK}= \frac{4 \mathcal{C}+\pi \log 2}{8\pi} \approx 0.232$, here $\mathcal{C} \approx 0.916$ is Catalan constant.

\subsection{D. Effective action for reparametrization modes in diffusive metal}
Inspired by the reparametrization symmetry in infrared, we redefine $\Sigma(\tau_1,\tau_2) \rightarrow \Sigma(\tau_1,\tau_2)+ \delta(\tau_1-\tau_2) \partial_{\tau_2} $, and bring the action to
\bea
	\frac{S_{UV}}{N} &=&\sum_x \frac12 \iint \Big[ \delta(\tau_{12}) (\partial_2 G_{\psi,x} + r\partial_2 G_{\eta,x})-\sqrt{r}t^{\prime2} ( G_{\eta,x} G_{\psi,x+1}- G_{\eta,x} G_{\psi,x})   \Big],  \\
	\frac{S_{IR}}{N} &=&\frac{1}{2} \sum_x -   [ tr \log(-\Sigma_{\psi,x}) + tr \log (-\Sigma_{\eta,x}) ] + \iint  \Big[ \Sigma_{\psi,x} G_{\psi,x} + r\Sigma_{\eta,x} G_{\eta,x} - \frac{J^2}{4} G_{\psi,x}^4 - \sqrt{r}(t^2+t^{\prime2}) G_{\psi,x} G_{\eta,x}  \Big]. \nn\\
\eea
The redefinitions effectively collect the time derivative term to $S_{UV}$. Moreover, since each unit cell decouples from others in $S_{IR}$, the saddle point equations given by $S_{IR}$ are exactly conformal invariant whose solutions are obtained above. Since $S_{IR}$ has reparametrization symmetry, it vanishes for reparametrization modes. However, this symmetry is explicitly broken by UV part, i.e., $S_{UV}$ will give a small action to reparametrization modes which dominants the four-point correlator in the infrared. The effective action for reparametrization modes is given by fluctuations around saddle point action, i.e.,
the effective action for reparametrization modes reads ${S[f]}=S_{UV}[\tilde{G}(f)]-S_{UV}[G(t)]$. For the first term in $S_{UV}$, we obtain the effective action using $\varepsilon$-expansion \cite{yoon2016, verlinde2017}, where $\varepsilon=\frac12-\Delta_\psi$:
\bea
	\frac{S^{(1)}_{UV}}{N} = \sum_x -\frac{1}{32\pi} (\frac{\sqrt{1-r}}{J}+ \frac{J}{t^2+t^{\prime2}} \sqrt{\frac{r^3}{1-r}})\int d\tau \{f_x, \tau\}.
\eea
where $\{f,\tau\}=\frac{f'''}{f'}-  \frac32 \Big( \frac{f''}{f'} \Big)^2$ denotes Schwartz derivative. For small reparametrization $f_x=\tau+\epsilon_x(\tau)$,
\bea
\frac{S^{(1)}_{UV}}{N} = \frac{1}{128\pi^2} (\frac{\sqrt{1-r}}{J}+ \frac{J}{t^2+t^{\prime2}} \sqrt{\frac{r^3}{1-r}}) \sum_{n,p} n^2(n^2-1)  \epsilon_{-n,-p} \epsilon_{n,p}
\eea
where $\epsilon_x(\tau)= \frac{1}{2\pi \sqrt{L}} \sum_{n,p}\epsilon_{n,p} e^{-in\tau+ipx}$, and we have set $\beta=2\pi$ for simplicity. For second term in $S_{UV}$, the reparametrization modes are defined as $\tilde{G}_{\psi,x}=G_\psi+\delta G_{\psi,x}, \tilde{G}_{\eta,x}=G_\eta+\delta G_{\eta,x}$. Though the saddle point solutions are uniform in spatial direction, their fluctuations are position-dependent \cite{gu2016}. Then the effective action reads
\bea
\frac{S^{(2)}_{UV}}{N}=\frac12 \sum_x \iint \sqrt{r} t^{\prime2} ( \delta G_{\eta,x} \delta G_{\psi,x}- \delta G_{\eta,x} \delta G_{\psi,x+1}),
\eea
where $\iint= \int d\tau_1 d\tau_2$. Explicitly, for conformal solutions $G(\tau)=\Lambda \frac{\text{sgn}(\tau)}{|2\sin\frac{\tau}{2}|^{2\Delta}}$, the reparametrization gives rise to
\bea
\delta G_x(\tau_1,\tau_2)= \frac{i\Delta}{\pi} G(\tau) \sum_n \epsilon_{n,x} h_n(\tau)e^{-in \bar{\tau}}.
\eea
where $h_n(\tau)\equiv \sin \frac{n\tau}{2} \cot \frac{\tau}{2}- n \cos \frac{n\tau}{2} $ and $\tau=\tau_1-\tau_2$, $\bar \tau= \frac12(\tau_1+ \tau_2)$. Then
\bea
\sum_x \iint (\delta G_{\eta,x} \delta G_{\psi,x}- \delta G_{\eta,x} \delta G_{\psi,x+1}) &=& \frac{\sqrt{r}}{64\pi (t^2+t^{\prime2})} \sum_{n,x} |n|(n^2-1) (\epsilon_{-n,x} \epsilon_{n,x} -\epsilon_{-n,x} \epsilon_{n,x+1}) \\
&= & \frac{\sqrt{r}}{64 \pi (t^2+t^{\prime2})} \sum_{n,p} |n|(n^2-1)(1-\cos p)\epsilon_{-n,-p} \epsilon_{n,p}
\eea
Plug these results in to $S^{(2)}$, we get
\bea
	\frac{S^{(2)}_{UV}}{N}=\sum_{n,p} \frac{1}{128\pi} \frac{r t^{\prime2}}{t^2+t^{\prime2}} (1-\cos p)|n|(n^2-1)\epsilon_{-n,-p} \epsilon_{n,p}
\eea
Finally, in terms of the infinitesimal modes, the effective action is given by
\bea
	\frac{S_{\text{eff}}}{N} &=& \frac{S^{(1)}_{UV}}{N}+\frac{S^{(2)}_{UV}}{N} = \frac{1}{2} \sum_{n,p} \Big[  \alpha_1 n^2(n^2-1)+ \alpha_2 2(1-\cos p)|n|(n^2-1) \Big] \epsilon_{-n,-p} \epsilon_{n,p},
\eea
where $\alpha_1=\frac{1}{64\pi^2} \Big(\frac{\sqrt{1-r}}{J}+ \frac{J}{t^2+t^{\prime2}} \sqrt{\frac{r^3}{1-r}}\Big)$, $\alpha_2=\frac{1}{128\pi} \frac{r t^{\prime2}}{t^2+t^{\prime2}}$. To restore the dimension, note that $\text{dim}[\epsilon_x(\tau)]=-1$, and $\text{dim}[\epsilon_n]=-2$, where $\text{dim}[...]$ denotes engineer dimension (not to confuse with scaling dimension), we obtain
\bea
	\frac{S_\text{eff}}{N}&=& \frac{\pi}{\beta}\sum_{n,p}  \Big[  \alpha_1 \omega_n^2 \Big( \omega_n^2-\big(\frac{2\pi}{\beta}\big)^2 \Big) + \alpha_2 p^2 |\omega_n| \Big( \omega_n^2- \big(\frac{2\pi}{\beta}\big)^2 \Big) \Big] \epsilon_{-\omega_n,-p} \epsilon_{\omega_n,p}.
\eea
where we have expanded for small momentum.

\subsection{E. Diffusive constant in diffusive metal}
According to Noether's theorem, the reparametrization modes in time direction couple to energy density, i.e., $\delta S= \int d\tau \epsilon \partial_\tau T$, where $\epsilon$ is reparametrization mode and $T$ is energy density. At zero momentum limit, the effective action is given by \bea
	\frac{S_\text{eff}}{N}= \frac{\alpha_1}{2} \int d\tau \big[(\epsilon'')^2-(\epsilon')^2 \big]=\frac{\alpha_1}{2} \int d\tau \epsilon \partial_\tau (\epsilon'''+ \epsilon'),
\eea
Then one finds that $T=\frac{\alpha_1 N}{2} (\epsilon'''+ \epsilon')$ or in frequency domain, $T_n=\frac{i\alpha_1 N}{4\pi} (n^3-n)\epsilon_n$. The correlation of energy density is given by
\bea
	\langle T_{-n} T_n \rangle=  \frac{N^2 \alpha_1^2 }{16\pi^2} (n^3-n)^2 \langle \epsilon_{-n} \epsilon_n \rangle.
\eea
For small momentum, one can directly generalize this function,
\bea
	\langle T_{-n,-p} T_{n,p} \rangle=\frac{N^2 \alpha_1^2}{16\pi^2}(n^3-n)^2 \langle \epsilon_{-n,-p} \epsilon_{n,p} \rangle
\eea
Plug the propagator from effective action into the correlation, we have
\bea
	\langle T_{-n,-p} T_{n,p} \rangle= \frac{N \alpha_1^2}{16\pi^2} \frac{n^2(n^2-1)^2}{\alpha_1 n^2(n^2-1)+ \alpha_2 p^2 |n|(n^2-1)}=\frac{N\alpha_1}{16\pi^2}\frac{|n|(n^2-1)}{|n|+ \frac{\alpha_2}{\alpha_1} p^2}.
\eea
Note that $\text{dim}[T(\tau)]=1$, we can also restore temperature:
\bea
	\langle T_{-\omega_n,-p} T_{\omega_n,p} \rangle=\frac{N \alpha_1}{16\pi^2}\frac{2\pi}{\beta} \frac{|\omega_n|\Big[ \big(\frac{\beta \omega_n}{2\pi} \big)^2-1\Big]}{|\omega_n|+ \frac{\alpha_2}{\alpha_1} p^2}.
\eea
Analytical continuation from upper half-plane, i.e. $i \omega_n \rightarrow \omega+ i\delta$, the retarded correlation function is
\bea
\langle T_{-\omega,-p} T_{\omega,p} \rangle= \frac{N \alpha_1}{8\pi\beta} \frac{Dp^2}{-i\omega+D p^2},
\eea
where $D\equiv \frac{\alpha_2}{\alpha_1}$ and we have implicitly extracted a contact term \cite{gu2016,sachdev2016}, i.e., $\langle T_{-\omega,-p} T_{\omega,p} \rangle- \langle T_{-\omega,0} T_{\omega,0} \rangle$.

\subsection{F. Chaos and butterfly velocity in diffusive metal}
Four-point functions are defined as
\bea
	F_{ab,xy}(\tau_1, \tau_2, \tau_3, \tau_4) &=& \frac{1}{N_a N_b} \sum_{i,j} \langle a_{i,x}(1) a_{i,x}(2) b_{j,y}(3) b_{j,y}(4) \rangle=  G_a(\tau_1, \tau_2)G_b(\tau_3, \tau_4)+ \frac{1}{N} \mathcal{F}_{ab,xy}(\tau_1, \tau_2, \tau_3, \tau_4),
\eea
where $\mathcal{F}_{ab,xy}(\tau_1, \tau_2, \tau_3, \tau_4) \equiv \langle \delta G_{a,x}(\tau_1, \tau_2) \delta G_{b,y}(\tau_3, \tau_4) \rangle$ and $a,b=\psi,\eta$. Using the translational symmetry,
\bea
	\mathcal{F}_{ab,xy}(\tau_1, \tau_2, \tau_3, \tau_4)=\frac{1}{L} \sum_p \mathcal{F}_{ab,p}(\tau_1, \tau_2, \tau_3, \tau_4) e^{ip(x-y)},
\eea
with $\mathcal{F}_{ab,p}(\tau_1, \tau_2, \tau_3, \tau_4)=L \langle \delta G_{a,-p}(\tau_1, \tau_2) \delta G_{b,p}(\tau_3, \tau_4) \rangle$. Thus
\bea
\frac{\mathcal{F}_{ab,p}(\tau_1, \tau_2, \tau_3, \tau_4)}{G_{a}(\tau_1, \tau_2) G_{b}(\tau_3, \tau_4)}= \frac{N\Delta_a\Delta_b}{\pi^2} \sum_{n} \langle \epsilon_{-n,-p} \epsilon_{n,p} \rangle h_n(x_{12}) h_n(x_{34}) e^{-in (y_{12}-y_{34})}.
\eea
For OTOC, let $\tau_1=\beta+it, \tau_4=\frac{3\beta}{4}, \tau_2= \frac{\beta}{2}+it, \tau_3=\frac{\beta}{4}$, then $x_{12}=\frac{\beta}{2},x_{34}=-\frac{\beta}{2},y_{12}=\frac{3\beta}{4}+it, y_{34}=\frac{\beta}{2}$, and set $\beta=2\pi$ for simplicity, we have
\bea
\frac{\mathcal{F}_{ab,p}(\tau_1, \tau_2, \tau_3, \tau_4)}{G_a(\pi) G_b(-\pi)} &=& \frac{N\Delta_a\Delta_b}{\pi^2} \sum_{n} \langle \epsilon_{-n,-p} \epsilon_{n,p} \rangle h_n(\pi) h_n(-\pi) e^{nt-i\frac{n\pi}{2}} \\
 &=& \frac{N\Delta_a\Delta_b}{\pi^2} \sum_n \langle \epsilon_{-n,-p} \epsilon_{n,p} \rangle n^2 (\cos\frac{n\pi}{2})^2 e^{nt-i\frac{n\pi}{2}} \\
&=& \frac{N\Delta_a\Delta_b}{\pi^2} \sum_{n\ge2,even} \langle \epsilon_{-n,-p} \epsilon_{n,p} \rangle (-1)^{\frac{n}{2}} n^2  (e^{nt}+e^{-nt}).
\eea
Plug the propagator obtained from effective action into above equation, we have
\bea
\frac{\mathcal{F}_{ab,p}(\tau_1, \tau_2, \tau_3, \tau_4)}{G_a(\pi) G_b(-\pi)}= \frac{2\Delta_a\Delta_b}{\pi^2} \sum_{n\ge2,even} \frac{(-1)^\frac{n}{2} n^2 \cosh nt }{\alpha_1 n^2 (n^2-1) + 2\alpha_2 (1-\cos p)|n|(n^2-1)}.
\eea
To evaluate above summation, consider the integral
\bea
K= \int_{-i\infty}^{i\infty} \frac{d\omega}{2\pi i} \frac{\pi}{2} \frac{1}{\sin \frac{\pi \omega}{2}}\frac{\omega^2 \cosh \omega t }{\alpha_1 \omega^2 (\omega^2-1) + \alpha_2 p^2 |\omega|(\omega^2-1)}.
\eea
Making a large semicircle contour to $w \rightarrow +\infty$ of complex plane \cite{gu2016}, one finds according to residue theorem
\bea
K =-\sum_{n\ge2,even} \frac{(-1)^{\frac{n}{2}} n^2 \cosh nt }{\alpha_1 n^2 (n^2-1) + \alpha_2 p^2 |n|(n^2-1)}- \frac{\pi}{4} \frac{\cosh t}{\alpha_1+ \alpha_2 p^2}.
\eea
or
\bea
\sum_{n\ge2,even} \frac{(-1)^{\frac{n}{2}} n^2 \cosh nt }{\alpha_1 n^2 (n^2-1) + \alpha_2 p^2 |n|(n^2-1)}=- \frac{\pi}{4} \frac{\cosh t}{\alpha_1+ \alpha_2 p^2}- K.
\eea
There is no exponential term in $K$ \cite{gu2016}, thus the only exponential growth part is
\bea
\frac{\mathcal{F}_{ab,p}(\tau_1, \tau_2, \tau_3, \tau_4)}{G_a(\pi) G_b(-\pi)}\ni -\frac{\Delta_a\Delta_b}{2\pi} \frac{\cosh t}{\alpha_1+ \alpha_2 p^2}.
\eea
Fourier transform to real space, then we have
\bea
	\frac{\mathcal{F}_{ab,xy}(\tau_1, \tau_2, \tau_3, \tau_4)}{G_a(\pi) G_b(-\pi)} &\ni& -\frac{\Delta_a\Delta_b}{2\pi} \frac{1}{L} \sum_p \frac{\cosh t}{\alpha_1+ \alpha_2 p^2} e^{ip(x-y)}=-\frac{\Delta_a\Delta_b}{2\pi} \int \frac{dp}{2\pi} \frac{\cosh t}{\alpha_1+ \alpha_2 p^2} e^{ip(x-y)} \\
	&=& -\frac{\Delta_a\Delta_b}{4\pi\sqrt{\alpha_1 \alpha_2}} e^{-\frac{|x-y|}{v_B}} \coth t \approx -\frac{\Delta_a\Delta_b}{4\pi\sqrt{\alpha_1 \alpha_2}} e^{t-\frac{|x-y|}{v_B}},
\eea
The exponent growth of OTOC is now given by (we have restored dimensions)
\bea
\frac{F_{ab,xy}}{G_a(\pi) G_b(-\pi)}\sim 1- \frac{1}{N} \frac{\Delta_a\Delta_b}{4\pi\sqrt{\alpha_1 \alpha_2}} \sqrt{\frac{\beta}{2\pi}} e^{\frac{2\pi}{\beta}(t-\frac{|x-y|}{v_B})}.
\eea
Since $x, y$ is dimensionless here (we set the lattice constant to 1), $[v_B]=1$ and the butterfly velocity is given by
\bea
	v_B^2=\frac{\alpha_2}{\alpha_1} \frac{2\pi}{\beta}= \frac{\pi r\sqrt{1-r} t^{\prime2}}{2[(1-r) \frac{t^2+t^{\prime2}}{J}+ r^{\frac32} J]}\frac{2\pi}{\beta}.
\eea
When approaching transition point, $r\rightarrow 1$, butterfly velocity vanishes indicating a MBL phase.

\subsection{G. Many-body localized phase}
Similar to the case of $r\ll 1$, we also make a translational invariant ansatz for $r\gg 1$, with which the saddle point equation can be approximated by
\bea
	&& G_{\psi}^{-1}=-i\omega - \Sigma_{\psi}, ~~~~~~G_{\eta}^{-1}= -i\omega- \Sigma_{\eta}, \\
	&& \Sigma_{\psi}= \sqrt{r} \tilde t^2 G_{\eta},~~~~~~~~~~ \Sigma_{\eta}= \tilde t^2 G_{\psi}/\sqrt{r},~~~
\eea
where ${\tilde t}^2\equiv t^2+t'^2$. The exact solutions of the above Schwinger-Dyson equations can be obtained:
\bea
	G_\eta\! =\! \frac{2}{\!-i\omega\!+\!i \frac{(r-1)\tilde t^2}{\sqrt{r}\omega}\!-\! i \text{sgn}(\omega) \sqrt{\frac{(r-1)^2\tilde t^4}{r \omega^2}\!+\! \frac{2(r+1)\tilde t^2}{\sqrt{r}}\!+\! \omega^2   }},~~~~~  \label{Geta}\\
	G_\psi\! =\! \frac{2}{\!-i\omega\!-i\! \frac{(r-1)\tilde t^2}{\sqrt{r}\omega}\!-\! i \text{sgn}(\omega) \sqrt{\frac{(r-1)^2\tilde t^4}{r \omega^2}\!+\! \frac{2(r+1)\tilde t^2}{\sqrt{r}}\!+\!\omega^2  }}. ~~~~~\label{Gpsi}
\eea
The self-energy part $\Sigma_a$ seems to dominate the propagator $G_a$ at low frequency or energy due to the terms $\propto\frac{1}{\omega}$ in the denominators of Eqs. (\ref{Geta}-\ref{Gpsi}). It is true for $\Sigma_\psi$; but for $\Sigma_\eta$ the $\frac{1}{\omega}$ terms are cancelled in the limit of low frequency.

\subsection{G. A modified model}
In order to lift the zero modes in MBL phase, we add random quadratic couplings into the Hamiltonian \cite{altman2017, song2017}. The modified Hamiltonian is
\bea
	H' = \sum_{x=1}^{L} \Big[ \frac{1}{4!} \sum_{ijkl} J_{ijkl,x} \psi_{i,x} \psi_{j,x} \psi_{k,x}\psi_{l,x}+  \frac12 \sum_{\alpha,\beta} V_{\alpha\beta,x} i\eta_{\alpha,x} \eta_{\beta,x }+ \sum_{i\alpha} \Big( t_{i\alpha,x} i\psi_{i,x} \eta_{\alpha,x} + t_{i\alpha,x}'  i\eta_{\alpha,x} \psi_{i,x+1} \Big) \Big],
\eea
where $J_{ijkl,x}, t_{i\alpha,x}, t'_{i\alpha,x}$ are random variables same as before, and $V_{\alpha\beta,x}$ refers to onsite random quadratic couplings satisfying Gaussian distribution with $\langle V_{\alpha\beta,x}\rangle=0$, $\langle V_{\alpha\beta,x}^2 \rangle=\frac{V^2}{M}$. We calculate the distribution of level-spacing ratios for the modified Hamiltonian as shown in Fig. \ref{distribution2}, from which one can infer that this system also has a transition from diffusive metal to MBL phase.

\begin{figure}[t]
\subfigure[]{\label{distr1}\includegraphics[width=5cm]{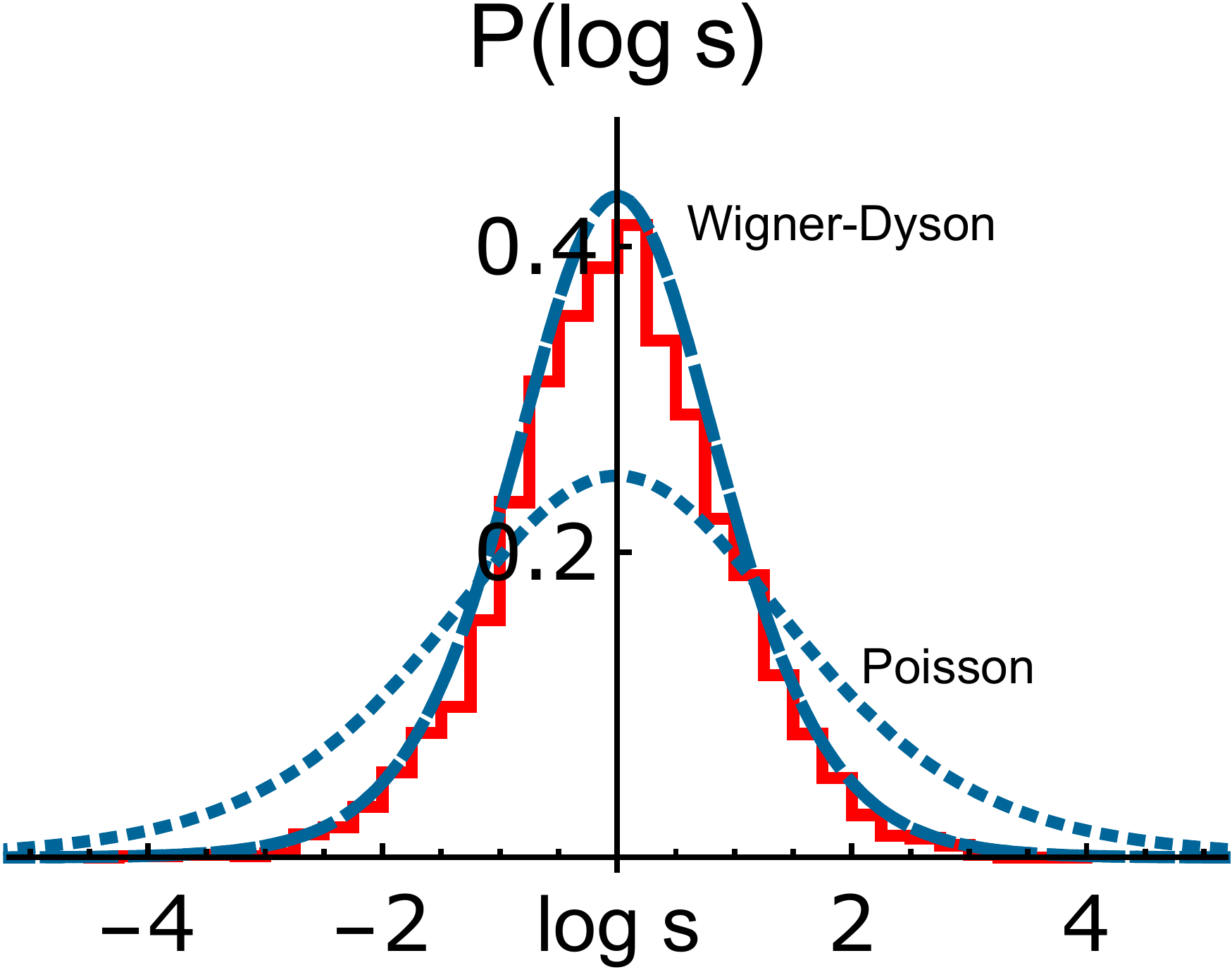}}
\subfigure[]{\label{distr2}\includegraphics[width=5cm]{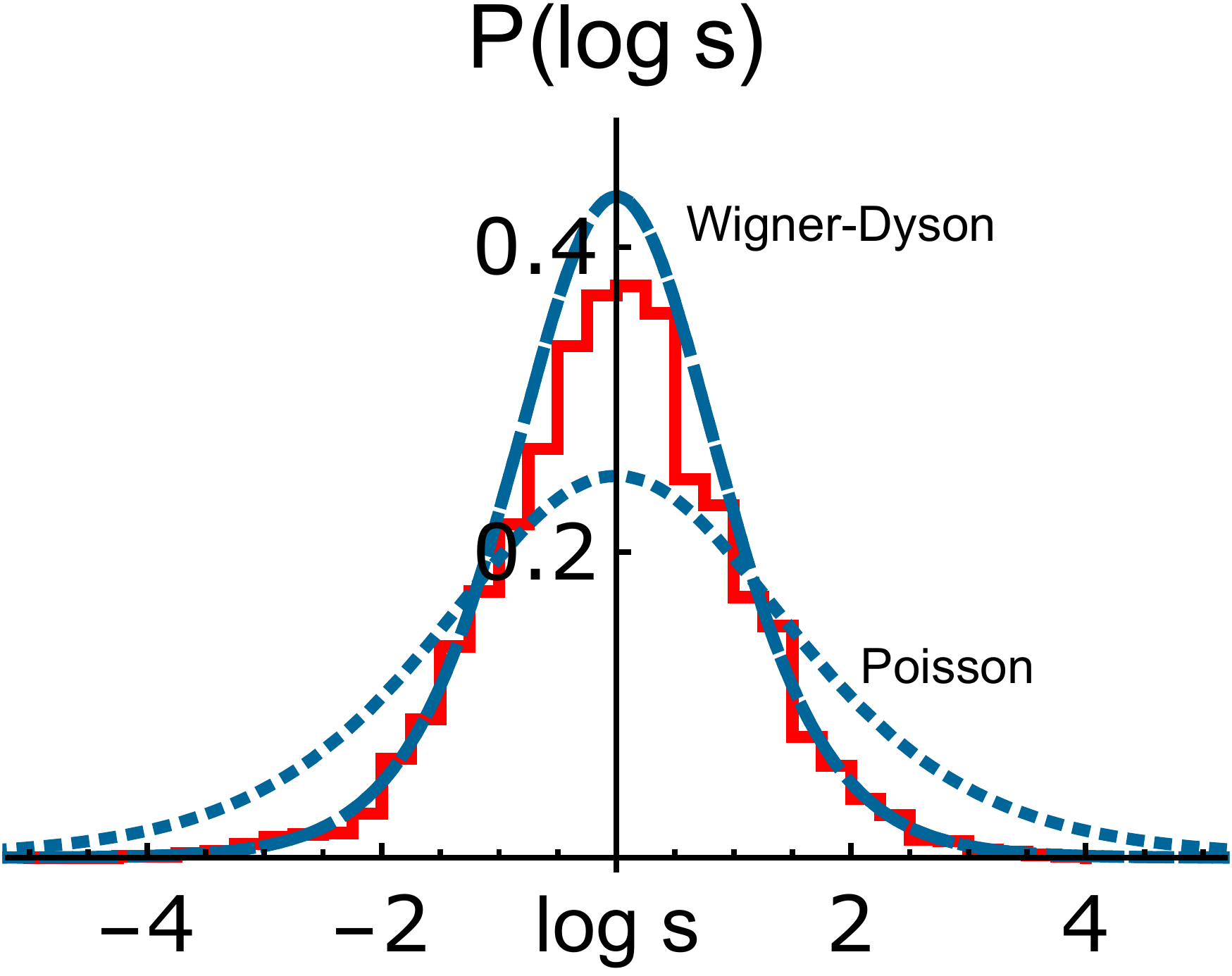}}
\subfigure[]{\label{distr3}\includegraphics[width=5cm]{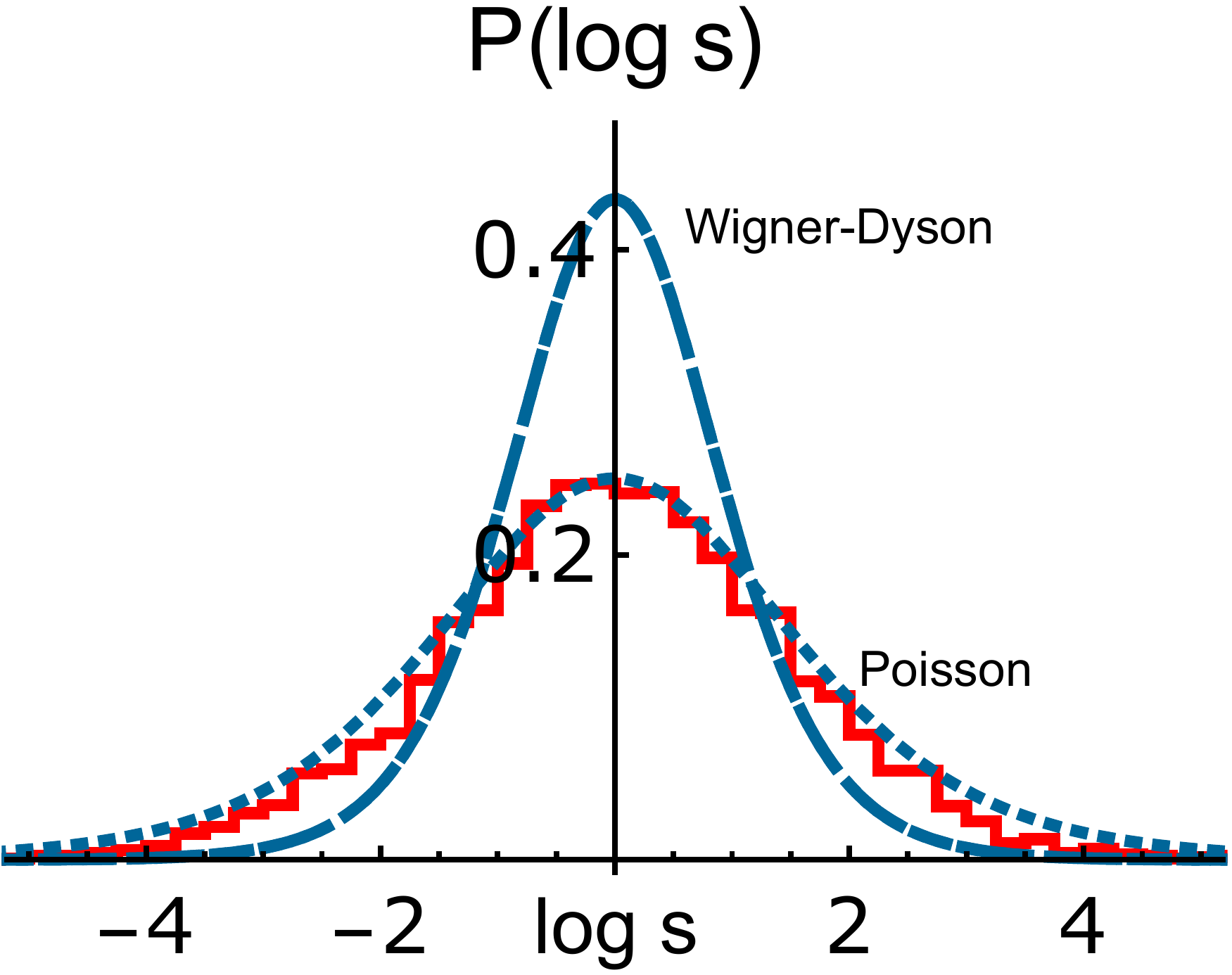}}
\caption{\label{distribution} The distribution of level-spacing ratios for the cases of $(N$,$M)$=(6,4), (5,5) and (4,6) are shown in (a), (b) and (c), respectively.  The results (red solid line) are obtained by exactly diagonalizing the generalized SYK model on the six-site chain with $N$+$M$=10 Majorana fermions in each unit cell and with $J=0.8$, $t=1.2$, $t'=0.1$, $V=0.2$. The Wigner-Dyson distribution (dashed line) implies thermalization while Poisson distribution (dotted line) implies MBL. \label{distribution2}}
\end{figure}
\end{widetext}

\begin{thebibliography}{99}
\bibitem{footnote2} For $N$=$\infty$, the random-hopping Majorana fermion model with $J$=$0$ would have a finite diffusive constant, which is actually an artifact of $N$=$\infty$. When $N$ is large but not infinity, the system with $r$$>$$1$ and $J$=$0$ would be Anderson localized.
\end{thebibliography}
\end{document}